\documentclass{article}
\usepackage{amsmath, amssymb, amsfonts}
\title{Comment on ''Black Hole with Quantum Potential''}  
\author{Hristu Culetu, \\Ovidius University, Dept.of Physics and Electronics, \\ Mamaia Avenue 124, 900527 Constanta, Romania, \\e-mail : hculetu@yahoo.com}

\begin{document}
\numberwithin{equation}{section}
\pagenumbering{arabic}
\maketitle
\newcommand{\fv}{\boldsymbol{f}}
\newcommand{\tv}{\boldsymbol{t}}
\newcommand{\gv}{\boldsymbol{g}}
\newcommand{\OV}{\boldsymbol{O}}
\newcommand{\wv}{\boldsymbol{w}}
\newcommand{\WV}{\boldsymbol{W}}
\newcommand{\NV}{\boldsymbol{N}}
\newcommand{\hv}{\boldsymbol{h}}
\newcommand{\yv}{\boldsymbol{y}}
\newcommand{\RE}{\textrm{Re}}
\newcommand{\IM}{\textrm{Im}}
\newcommand{\rot}{\textrm{rot}}
\newcommand{\dv}{\boldsymbol{d}}
\newcommand{\grad}{\textrm{grad}}
\newcommand{\Tr}{\textrm{Tr}}
\newcommand{\ua}{\uparrow}
\newcommand{\da}{\downarrow}
\newcommand{\ct}{\textrm{const}}
\newcommand{\xv}{\boldsymbol{x}}
\newcommand{\mv}{\boldsymbol{m}}
\newcommand{\rv}{\boldsymbol{r}}
\newcommand{\kv}{\boldsymbol{k}}
\newcommand{\VE}{\boldsymbol{V}}
\newcommand{\sv}{\boldsymbol{s}}
\newcommand{\RV}{\boldsymbol{R}}
\newcommand{\pv}{\boldsymbol{p}}
\newcommand{\PV}{\boldsymbol{P}}
\newcommand{\EV}{\boldsymbol{E}}
\newcommand{\DV}{\boldsymbol{D}}
\newcommand{\BV}{\boldsymbol{B}}
\newcommand{\HV}{\boldsymbol{H}}
\newcommand{\MV}{\boldsymbol{M}}
\newcommand{\be}{\begin{equation}}
\newcommand{\ee}{\end{equation}}
\newcommand{\ba}{\begin{eqnarray}}
\newcommand{\ea}{\end{eqnarray}}
\newcommand{\bq}{\begin{eqnarray*}}
\newcommand{\eq}{\end{eqnarray*}}
\newcommand{\pa}{\partial}
\newcommand{\f}{\frac}
\newcommand{\FV}{\boldsymbol{F}}
\newcommand{\ve}{\boldsymbol{v}}
\newcommand{\AV}{\boldsymbol{A}}
\newcommand{\jv}{\boldsymbol{j}}
\newcommand{\LV}{\boldsymbol{L}}
\newcommand{\SV}{\boldsymbol{S}}
\newcommand{\av}{\boldsymbol{a}}
\newcommand{\qv}{\boldsymbol{q}}
\newcommand{\QV}{\boldsymbol{Q}}
\newcommand{\ev}{\boldsymbol{e}}
\newcommand{\uv}{\boldsymbol{u}}
\newcommand{\KV}{\boldsymbol{K}}
\newcommand{\ro}{\boldsymbol{\rho}}
\newcommand{\si}{\boldsymbol{\sigma}}
\newcommand{\thv}{\boldsymbol{\theta}}
\newcommand{\bv}{\boldsymbol{b}}
\newcommand{\JV}{\boldsymbol{J}}
\newcommand{\nv}{\boldsymbol{n}}
\newcommand{\lv}{\boldsymbol{l}}
\newcommand{\om}{\boldsymbol{\omega}}
\newcommand{\Om}{\boldsymbol{\Omega}}
\newcommand{\Piv}{\boldsymbol{\Pi}}
\newcommand{\UV}{\boldsymbol{U}}
\newcommand{\iv}{\boldsymbol{i}}
\newcommand{\nuv}{\boldsymbol{\nu}}
\newcommand{\muv}{\boldsymbol{\mu}}
\newcommand{\lm}{\boldsymbol{\lambda}}
\newcommand{\Lm}{\boldsymbol{\Lambda}}
\newcommand{\opsi}{\overline{\psi}}
\renewcommand{\tan}{\textrm{tg}}
\renewcommand{\cot}{\textrm{ctg}}
\renewcommand{\sinh}{\textrm{sh}}
\renewcommand{\cosh}{\textrm{ch}}
\renewcommand{\tanh}{\textrm{th}}
\renewcommand{\coth}{\textrm{cth}}

\begin{abstract}
Few comments upon Ali and Khalil paper \cite{AK} are pointed out. Their modified Schwarzschild metric seems not to be new as it has the same structure as Eq. (2.6) from Ref.4. Their black hole temperature $T$ and heat capacity $C$ correspond exactly to the Reissner-Nordtstrom values, with $\hbar \eta$ instead of $Q^{2}$. The expression for the black hole entropy turns out to be erronous and does not fit with the other authors' calculations. Moreover, a lot of equations have wrong physical units (terms in the same equation have different units).
\end{abstract}

 Black holes are ideal laboratories to investigate Quantum Gravity. The singularities which were predicted to form inside them are generally regarded as indicating the breakdown of General Relativity (GR), requiring quantum corrections. Due to those quantum corrections, a remnant may form so that the classical singularity at $r = 0$ is removed. 

Das \cite{SD} proposed a new semiclassical model for QG by replacing the classical geodesics with quantum ones, leading to corrections to the Raychaudhuri equation, with a Bohmian-type quantum potential. Recently, Ali and Khalil \cite{AK} introduced quantum corrections in black hole (BH) physics using a similar approach as the author of \cite{SD}. They derived the quantum Raychaudhuri equation for null geodesics and, from here, a modified version of the Schwarzschild (KS) line element with a term in the metric depending on $\hbar$. 

We notice firstly that, at p.3 (Eqs.5 and 8), the authors of \cite{AK} should have specified the significance of $f_{ab}$ and $\sigma^{ab}$ - the fact that $f_{ab}$ represents an antisymmetric matrix and $\sigma^{ab}$ is related to the Dirac matrices \cite{SD}. In addition, in the Introduction they rectified the expression of the induced metric $h_{ab}$, probably due to a suggestion from \cite{HC}, a paper which, however, has not been cited by the authors of \cite{AK}. 

After these formal observations, let us comment on the quantum-modified KS metric. Even though authors' equation (17) for the scalar expansion leads to $\Theta = 2/r$ when $\alpha(r) \beta(r) = 1$, one turns out that it is not generally valid. Indeed, from their Eqs. 4 and 19 and from $N_{a}N^{a} = 0,~k_{a}N^{a} = -1$ and $k_{a}k^{a} = 0$, one finds that
 \begin{equation}
k^{a} = \left(\frac{1 + \sqrt{\alpha \beta}}{2 \alpha}, \frac{1 + \sqrt{\alpha \beta}}{2 \sqrt{\alpha \beta}}, 0, 0 \right),
 \label{0.1}
 \end{equation}  
with $a = (t, r, \theta, \phi)$. Eq. (0.1) yields the expansion 
 \begin{equation}
\Theta = \nabla_{a}k^{a} = \frac{4\alpha \beta + 4\sqrt{\alpha \beta} + r\alpha' \beta + r\alpha \beta'}{4r\alpha \beta},
 \label{0.2}
 \end{equation}  
where the prime stands for the derivative with respect to $r$. We cannot get rid completely of the term $ 4 \sqrt{\alpha \beta}$ from the numerator because $k^{r}$ contains it at the denominator. Moreover, (0.2) reduces also to $2/r$ when $\alpha \beta = 1$. In addition, with, for example, $\alpha,\beta = const.$ but $\alpha \beta \neq 1$, their expression (17) gives $\Theta = 2/r$ but Eq. (0.2) leads to $\Theta = (1 + \sqrt{\alpha \beta})/(r \sqrt{\alpha \beta})$, namely our $\Theta$ depends on $\alpha$ and $\beta$, as it should be, even though the authors' geometry (12) becomes flat. That is an extra evidence on the correctness of our expression (0.2).

 When Ali and Khalil correctly obtained the differential equation (20), we think they should have indicated that $h^{ab}$ from the reduced Raychaudhuri equation 
  \begin{equation}
	h^{ab} \nabla_{a}\nabla_{b}R = 0
 \label{0.3}
 \end{equation}  
is two-dimensional, with $h^{tt} = h^{rr} = h^{rt} = 0$ ($R$ is the scalar curvature for their metric (12), with $\beta = 1/\alpha$). 

Although the authors have not specified the sign of the constant $\eta$ in Eq. 22, we suppose their choice was $\eta > 0$, as could be seen from the comparison with the Reissner-Nordtstrom (RN) black hole at the top of the page 8: $\eta \hbar \rightarrow Q^{2}$. One notices that the line element (22) is a particular case of the metric (2.2) from \cite{HC1}, with $k = 2/e$ and $lne = 1$. Indeed, for $r >> 2/em$, $f(r)$ from (2.5) of \cite{HC1} acquires the form 
  \begin{equation}
   f(r) \approx 1 - \frac{2m}{r} + \frac{4l_{P}^{2}}{er^{2}}.	
 \label{0.4}
 \end{equation}  
One sees that $(-g_{tt})$ from Eq. 22 and $f(r)$ from (0.4) coincide when $\eta = 4/e$. This may also be observed from the expression of the stress tensor. The same approximation leads to a $T^{a}_{~b}$ (Eq. 3.1 from \cite{HC1}) identical with that from Eq. 26 \cite{AK}. However, the exact form of $T^{a}_{~b}$ from \cite{HC1} is nonsingular at $r = 0$ compared with $T^{a}_{~b}$ from \cite{AK} which diverges at the origin. 

Using Painleve-Gullstrand (PG) coordinates, Ali and Khalil argued that the quantal trajectories never feel the singularity at $r = 0$ : their form (30) of the PG line element turns out to be new, at least as far as we know. The $r_{min} = \hbar \eta /2M$ introduces a constraint on the values of the radial coordinates as if a repulsive core sit at the origin. What the authors of \cite{AK} did not specify is that $(dr/dT)_{min} = 1 - M/\sqrt{\hbar \eta}$ from Fig.2 (dashed curve for outgoing light) might be positive when $M < \sqrt{\hbar \eta}$, or equivalently $M < M_{P}$, if $\eta$ were of the order of unity ($M_{P}$ is the Planck mass). 

Let us pass now to the comments regarding the thermodynamics of the BH (22). Authors' discussion concerning the temperature \footnote{The temperature $T$ from (35) is the same as that from the first version of the their preprint, only the algebraic form is different. A similar conclusion concerns the expression of the heat capacity $C$ from Eq. 37.} (35) in the extremal situation resembles a similar one, below Eq. (2.4) of Ref.4. Indeed, Ali and Khalil's mass $M = \sqrt{\hbar \eta}$ (with $\eta \approx 1$) corresponds to $m = m_{P}$ case from \cite{HC1} when the BH has a degenerate horizon (''frozen'' horizon \cite{BR}) and a soliton-like remnant is generated. Two observations on the expression (36) for the entropy $S$ are in order here. 

i)  In our opinion, the result (36) is invalid. Calculating $S$ by means of the expression (35) for $T$, one indeed obtains
  \begin{equation}
	S = \int{\frac{dM}{T(M)}} = \pi r_{+}^{2} = \frac{A_{+}}{4},
 \label{0.5}
 \end{equation} 
where $r_{+} = M + \sqrt{M^{2} - \hbar \eta}$ is the event horizon radius and $A_{+}$ is its area. It is worth noting here that Eq. 37 from the first version 1509.02495v1 is missing in the revised one. Most likely the authors became aware it is erronous, as it was already suggested in \cite{HC} which, again, has not been cited by the authors of \cite{AK}. Of course, Eq. 37 from the first version is erronous because Eq. 36 for the entropy $S$ is also incorrect (in both original and revised versions).

ii) As even Ali and Khalil have noticed, their metric (22) is a RN-type metric, with $\hbar \eta \rightarrow Q^{2}$. So, their temperature is exactly RN BH temperature, when $Q^{2}$ is replaced with $ \hbar \eta$. Therefore, our result (0.5) is not a surprise as it is exactly the RN BH entropy.

We indeed cannot apply (0.5) for the extremal case because the temperature is vanishing. However, we may use Padmanabhan prescription \cite{TP} (see also \cite{HC1, HC2})
  \begin{equation}
	S = \left(\frac{|W|}{2T}\right)_{H},
 \label{0.6}
 \end{equation} 
where $W$ is the Komar mass (see \cite{HC1, HC2} and Refs. therein) 
  \begin{equation}
  W = 2TS,~~~with~~ dW = TdS - p_{r}dV,
 \label{0.7}
 \end{equation} 
where ''H'' signifies the value on the horizon and $p_{r}$ is the radial pressure. Although $W = M - (\hbar \eta/r)$ and $T = \kappa/2\pi$ are zero at the horizon $r = r_{+} = M$ ($\kappa = (Mr_{+} - \hbar \eta)/r_{+}^{3}$ is the surface gravity), their ratio is finite and we get exactly $S = \pi r_{+}^{2}$. 

 Let us observe that the correct expression (37) for the heat capacity $C$ is obtained not from authors' equations (35) and (36) but from (35) and Eq. (0.5) of the present paper.

Regarding the time-dependent modified metric, it is not clear for us its physical significance. Even the authors of \cite{AK} admit some unphysical features of Eq. 45. In addition, it is not clear what are the units of the spread $\sigma$ of the wave pocket. From their Eq. (38) it should have dimension $(length)^{2}$ but from (41) and(42) results that $\sigma$ is a distance, as it should be. Similar arguments may be expressed for $f(r,t)$: in Eq.(40) it appears to be a $(length)^{-2}$. However, that seems to be in contradiction with (44), where its r.h.s. has terms with the units $(length)^{-4}$. Even Eq. (45) contains terms on its r.h.s. with different dimensions. 

As far as the physical constraints on $\eta$ are concerned (Eqs. 52 and 59), they are of little use due to their huge values. Indeed, the authors' Eqs. (34) - (37) require $M^{2} \geq \eta\hbar$. For, say, $\eta = 10^{70}$, one obtains $M = 10^{35}M_{P} \approx 10^{27} Kg$, which is not a reasonable constraint (a remnant cannot have such a large mass). Unfortunately, the authors of \cite{AK} did not give an order of magnitude of the constant $\eta$.

To summarize, few comments upon Ali and Khalil article were pointed out in this paper. Their modified KS metric (22) seems not to be new as it has the same structure as (2.6) from Ref.4. The BH temperature (35) and the heat capacity (37) correspond exactly to the RN values, with $\hbar \eta$ instead of $Q^{2}$. But the expression (36) for the entropy turn out to be incorrect for the reasons already given above. It is worth noting that Lashin \cite{EL} (see also \cite{TDDH}) expressed doubts on the correctness of cosmology from quantum potential and showed that the approach of using a quntum corrected Raychaudhuri equation is unsatisfactory because it cannot predict dynamics, being a kinematical equation.

\end{document}